\documentstyle[newarcrc,fleqn,psfig]{article}
\title{RXTE Observations of Cygnus X-3}

\author{M. L. McCollough\address{Universities Space Research Association,
Huntsville, AL 35806, U.S.A.}, C. R. Robinson$^{\rm a}$, S. N. Zhang$^{\rm a}$, %
      B. A. Harmon \address{ES84 NASA/Marshall Space Flight Center, Huntsville, AL
      35812, U.S.A.}, %
      W. S. Paciesas\address{University of Alabama in Huntsville, Huntsville, AL
      35899, U.S.A.}, \\ S. W. Dieters$^{\rm c}$, %
      R. M. Hjellming\address{National Radio Astronomy Observatory/VLA, Socorro,
      NM 87901, U.S.A. }, M. Rupen$^{\rm d}$, %
      A. J. Mioduszewski \address{JIVE/National Radio Astronomy Observatory/VLA, Socorro,
      NM 87901, U.S.A. }, %
      E. B. Waltman \address{Naval Research Laboratory, Washington, D.C. 20375, 
      U.S.A. }, %
      \\ F. D. Ghigo \address{National Radio Astronomy Observatory/GBI, Green Bank,
      WV 24944, U.S.A.}, %
      G. G. Pooley \address{Mullard Radio Astronomy Observatory, Cambridge, U.K.},%
      R. P. Fender \address{University of Amsterdam, Kruislaan 403, 1098 SJ 
      Amsterdam, The Netherlands}, %
      W. Cui \address{Massachusetts Institute of Technology, Cambridge, MA 
      02139, U.S.A. }, %
      S. Trushkin \address{Special Astrophysical Observatory, Nizhnij Arkhyz, 
      357147, Russia}}

\begin{document}
\maketitle

\begin{abstract}
In the period between May 1997 and August 1997 a series of pointed RXTE
observations were made of Cyg X-3.  During this period Cyg X-3 made a transition
from a quiescent radio state to a flare state (including a major flare) and then
returned to a quiescent radio state.  Analyses of the observations
are made in the context of concurrent observations in the hard X-ray 
(CGRO/BATSE), soft X-ray (RXTE/ASM) and the radio (Green Bank Interferometer,  
Ryle Telescope, and RATAN-600).  Preliminary analyses of the observations are 
presented.
\end{abstract}

\section{Introduction}

     Cyg X-3 represents one of the most unusual X-ray binaries to have ever 
been observed (for a review see \cite{bb_c 1988}).  It does not 
fit well into any of the established classes of X-ray 
binaries.  Its orbital period, 4.8 hours, is typical of a 
low mass X-ray binary, but infrared observations 
\cite{van_ker 1992}
indicate that the 
mass donating star may be a Wolf-Rayet star, which would make the system a high 
mass X-ray binary.  In addition, Cyg X-3 undergoes giant radio 
outbursts and there is evidence of jet-like structures moving away from 
Cyg X-3 at 0.3--0.9\,$c$ [Gregory et al. 1972, Geldzahler et al. 1983, Mioduszewski
et al. 1998a, 1998b].

     Previous X-ray observations of Cyg X-3 have shown a strong 4.8 hour 
modulation associated with the orbital period.  Modulations with the same 
period have been confirmed in the infrared \cite{Beck 1972,Mason 1986}.  
  From Ginga observations 
\cite{Wat 1994}, it appears that 
the giant radio flares occur only when the soft X-ray flux is high.
Slow (0.02 -- 0.001 Hz) QPOs have occasionally been seen 
in the X-ray \cite{van_jan 1985}.

\subsection{Recent Discoveries}

     In recent studies [McCollough et al. 1997a, 1997b, 1998a] the 
20--100 keV hard X-ray (HXR) emission, detected from Cyg X-3 by CGRO/BATSE, was
compared with the radio data from the Green Bank Interferometer (GBI), 
radio data from Ryle Telescope, and the 2--12 keV soft X-ray (SXR) flux detected 
by the ASM
on RXTE. These comparisons show that:
{\it (a)} During times of quiescent radio emission (moderate radio brightness of
$\sim 100$ mJy with low 
variability) the HXR flux anticorrelates with the radio.  It is during this
time that the HXR reaches its highest level. (see Fig. 1)
{\it (b)} During periods of flaring activity in the radio the HXR flux switches 
from an anticorrelation to a correlation with the radio.  In particular, for
major radio flares and the quenched radio emission (very low radio fluxes of 
10--20 mJy) that precedes these flares, the correlation is strong. (see Fig. 1)
{\it (c)} The HXR flux has been shown to anticorrelate with the SXR flux.  This 
occurs in
both the low and high SXR states. 
{\it (d)} Watanabe's [1994] results that the flaring periods in
the radio occur during the high SXR states has been confirmed.

     For the entire BATSE light curve of Cyg X-3 (over 
2200 days) a comparison with the radio and SXR data show that the above
discoveries hold over all of the overlaps between the data sets.  It has also been
shown that the onset of large radio flares ($>$ 5 Jy) can be predicted by noting
when the HXR drops below the BATSE detection limit for several days.

In February 1997 a large radio flare (10 Jy) was observed by GBI and Ryle after
a period of quenched emission.     
The flare triggered VLBA observations to obtain high resolution radio images of
Cyg~X-3 during the major flare.  The images showed a striking 
one-sided jet.   For a distance of 10 kpc using a twin-jet model,
Mioduszewski et al. [1998a, 1998b] find a jet with a velocity of 0.9\,$c$ and  
an inclination
of $2^{\circ}$ to our line of slight.  Thus Cyg X-3 appears to be similar to 
the superluminal sources GROJ1655$-$40 and GRS1915+105.

\section{RXTE Observations}

To probe the various X-ray/radio states of activity in Cyg X-3 a series of 
target of opportunity
(ToO) observations were made with RXTE.  The ToO (a series of ten 6 ksec 
observations) was designed to start during an extended quenched radio
state and follow Cyg X-3 through a large flare and into radio quiescent state. 
The times of the various RXTE observations are indicated on Fig. 1 along with
HXR and radio measurements for this time period.  There
observations cover the time period from June--August 1997.  During this 
timespan radio observations were also made with the Ryle and RATAN-600 radio 
telescopes \cite{trush 1998}.  
Some preliminary
results are:

\begin{figure}[htb]
\begin{minipage}[t]{3.0in}
\psfig{file=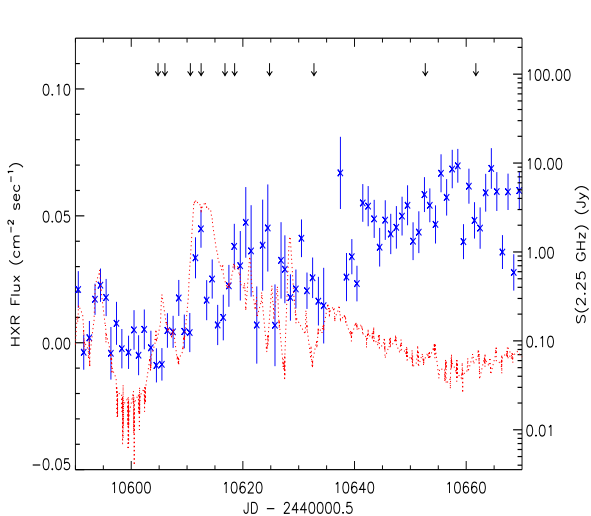,angle=90.0,height=2.75in,width=3.0in}
\caption{A plot of the daily BATSE 20--100 keV flux measurements of Cyg X-3 
(X with error
bars) for the time interval 10590--10670.  A power law with a fixed spectral 
index of --3 was used to
determine the HXR flux.  Overlayed on a log scale are the 2.25 GHz fluxes 
(dotted
line) measured by the GBI during the same time interval.  The arrows 
across the top of the plot indicate the times of pointed RXTE 
observations.}
\end{minipage}
\hspace{\fill}
\begin{minipage}[t]{3.0in}
\psfig{file=obs01_lc.ps,angle=-90.0,height=2.75in,width=3.0in}
\caption{A PCA lightcurve (all detectors, all channels) of the first Cyg X-3 
observation.  This observation was made
a few days after the radio came out of a quenched state.  Note the $\sim$ 15 \%
dip and recovery in the light curve near the beginning of the observation and 
the drop
in the count rate as a result of the 4.8 hr orbital modulation.}
\end{minipage}
\end{figure}

\subsection{Lightcurves}

All of the RXTE Proportional Counter Array (PCA) lightcurves show the 4.8 hr
orbital modulation.  In the first observation of the ToO, taken shortly after 
the radio came
out of an extended quenched state (see Fig. 2), the lightcurve showed large 
scale
fluctuations near the peak of the lightcurve and a smooth decline into minimum. 
This behavior is typical of the observations before the radio flare. 
In all other observations after the major flare, including a minor radio flare
period and radio quiescent period, there were large scale fluctuations throughout
the 4.8 hr cycle.

\subsection{Temporal Variations}

Preliminary temporal analysis of the RXTE data has not shown significant
features in power spectra in any of the observations (see Fig 3).  Scattering 
in the wind from the Wolf-Rayet companion can be expected to strongly suppress 
rapid X-ray variability by smearing, which explains the lack 
of short
time scale variability.  This lack of short timescale variability has also been
noted by \cite{Kita 1992} and \cite{Berg_van 1996}.  Modeling of
the cut-off frequency of the variability is underway to estimate wind
parameters.

\subsection{Hardness Ratios}

Hardness ratios were created from the ASM and BATSE data \cite{McC 1998b}.  
Hardness ratios from the ASM data alone show that during the radio quiescent 
periods the spectrum becomes
harder.  Coupled with an increase in the HXR seen concurrently in BATSE, this
indicates that the spectrum is exhibiting a pivoting 
behavior.   However, BATSE hardness ratios indicate that the spectrum above 20 keV 
becomes softer during
times of radio quiescence, which implies something more complicated than simple
spectal pivoting.  This
can be understood from examining the high energy spectra from RXTE (see below).

\subsection{Spectra}

Cyg X-3's X-ray spectrum is known to be complicated and several
components are needed to describe the spectrum \cite{Nak 1993}.
The most prominate components are an absorbed power-law with an 
exponential cut-off at high energy (a Comptonized spectrum) and a 
broad iron emission line.

In Fig. 4 we show two count spectra overlayed.  One is from just after a quenched
radio state and the other from radio quiescent state.  It can be seen
that the Cyg X-3 X-ray spectrum substantially changes for the different 
radio states in a way consistant with both ASM and BATSE hardness ratios.  Both 
spectra can be fit 
with the Nakamura et al. [1993] unified model.  In all cases the Comptonized
component dominates the spectrum.  

Spectra for other observations during the flaring activity are similar to the 
spectrum of the first observation, but show a pivoting of the spectrum around 
10 keV similar to what has been
observed in Cyg X-1 \cite{Zhang 1997}.

\begin{figure}[htb]
\begin{minipage}[t]{3.0in}
\psfig{file=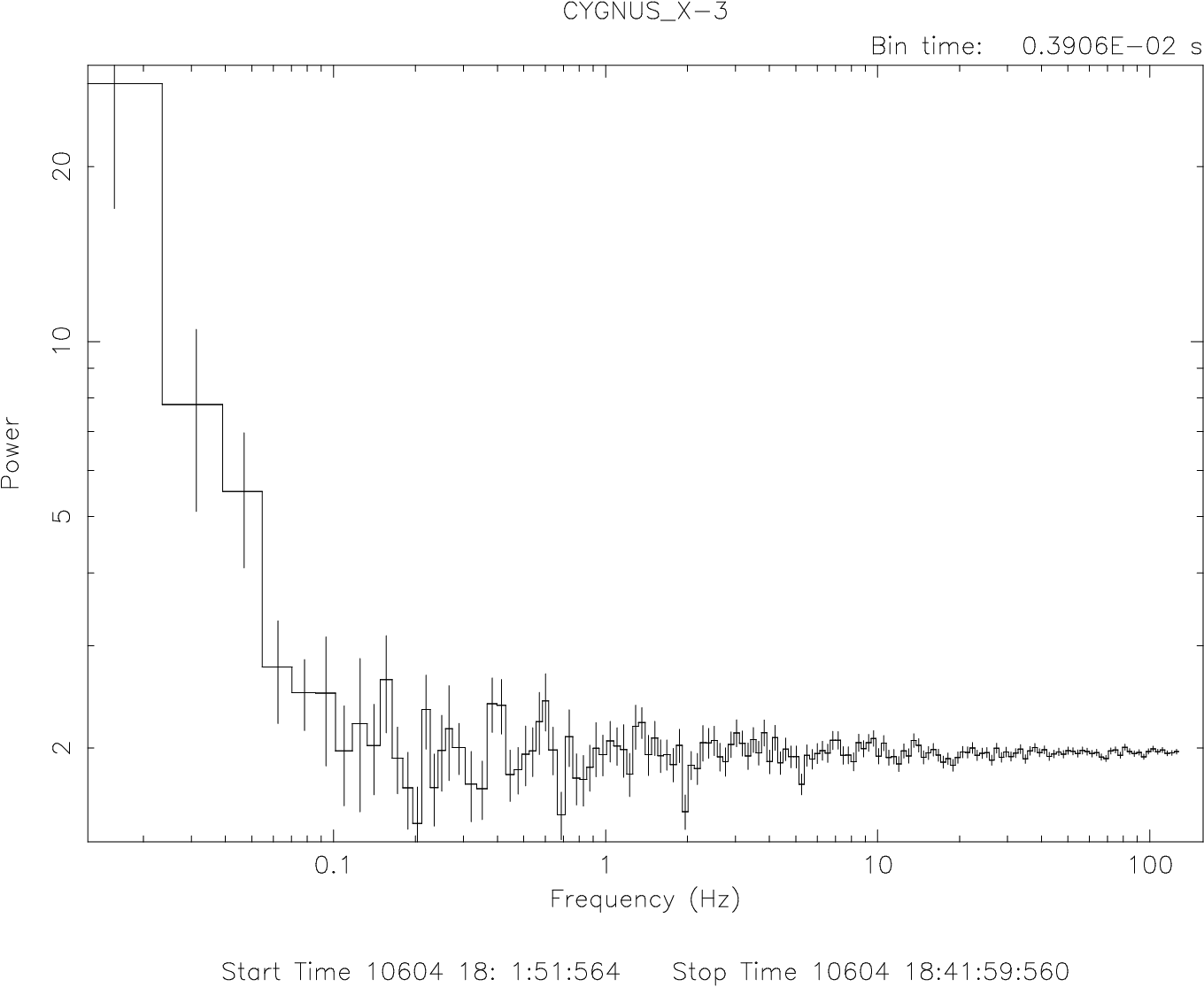,angle=-90.0,height=2.35in,width=3.0in}
\caption{A power spectrum of the initial half of observation 1.  No significant
features have been found.  The low frequency cutoff is being examined to
estimate wind parameters.}
\end{minipage}
\hspace{\fill}
\begin{minipage}[t]{3.0in}
\psfig{file=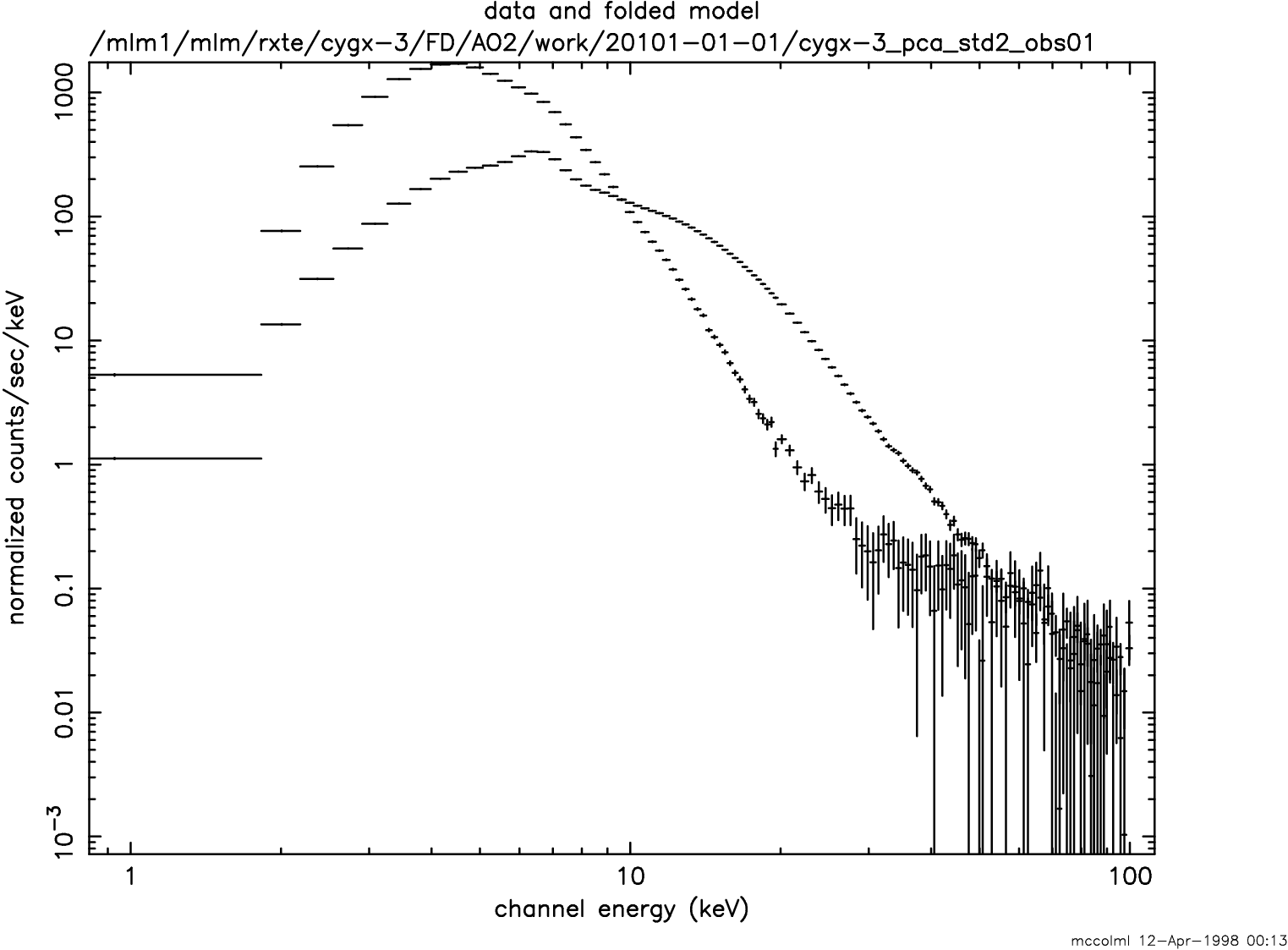,angle=-90.0,height=2.35in,width=3.0in}
\caption{Count spectra of the 1st and 9th observations overlayed to contrast the
change in spectra with the change in state.  In the 9th observation the Fe line
at 6.7 keV is more prominent.} 
\end{minipage}
\end{figure}

\section{Summary}

Preliminary analysis of the pointed RXTE observations has shown distinct
differences between lightcurves and spectra that depend upon the activity in 
the radio.   
Temporal analysis of the data has yet to show any short timescale variability. 

\end{document}